# An ultra-broadband continuously-tunable polarization entangled photon pair source covering the C+L telecom bands based on a single type-II PPKTP crystal

Zhi-Yuan Zhou[1+], Yun-Kun Jiang[2+#], Dong-Sheng Ding[1], Bao-Sen Shi[1*]

[1]*Key Laboratory of Quantum Information, University of Science and Technology of China, Hefei 230026, P. R. China*

[2]*College of Physics and Information Engineering, Fuzhou University, Fuzhou, 350002, P. R. China*

[+]These authors contribute equally to this work.

[*] *drshi@ustc.edu.cn*

[#] *ykjiang@fzu.edu.cn*

**Abstract.** In this work, we report on the preparation of an ultra-broadband continuously-tunable, highly polarization-entangled photon-pair source covering the C+L telecom bands using a single type-II phase-matched bulk PPKTP crystal. Over 60 nm tuning band of the down-converted photons is achieved experimentally. The broad tuning ability is due to operating near an extended phase match point in this crystal. The S parameter value of CHSH inequality is in the range between 2.60±0.04 (minimum) and 2.72±0.07 (maximum), which clearly demonstrates the high entanglement of the source over the whole tunable range. This is the first realization of such a broadband tunable degenerate polarization entanglement pair source in the telecom band. Combined with dense wavelength-division multiplexing technique, our source could be used to improve the transmission capacity of a single communication channel in the field of quantum communication.

**OCIS codes:** (190.4410) Nonlinear optics, parametric process; (270.5290) Photon statistics; (060.0060) Fiber optics and optical communications; (040.1345) Photon counting; (270.5565) Quantum communication.

## 1. Introduction

Quantum entanglement is an important resource in quantum communication applications such as quantum key distribution [1], quantum teleportation [2, 3], entanglement swapping [4] and quantum repeaters [5-7]. Quantum bits of information (qubits) can be encoded on different degrees of photons, the most common being used are polarization and time-bin. In experiments, spontaneous parametric down-conversion (SPDC) in nonlinear bulk crystals [8-14], or waveguide crystals [15-18], and spontaneously Raman scattering or four-wave mixing in atomic ensembles [19-26] have been widely employed to produce polarization [8, 9, 11-14, 16] or time-bin [15, 17, 18] entangled photon-pairs in either the visible [8-14, 20-23] or the telecom [15-19, 24, 25] band.

However, the emergence of long distance quantum communication [27, 28] demands a high quality entangled photon-pair source in telecom band due to low transmission loss in the fiber. In addition, the bandwidth of the photons should be narrow enough to avoid both chromatic dispersion and polarization mode dispersion [29] in the fiber. So far all photon-pair sources reported are centered at a single wavelength, the wavelength tuning ability of them is very limited. In traditional optical communication networks, wavelength-division multiplexing (WDM) is a common used technique to burst the transmission capacity of a single communication channel. To apply WDM technique, one needs a broadband light source. Here we report on, for the first time, the preparation of an ultra-broadband continuously-tunable polarization-entangled photon-pair source using a single bulk PPKTP crystal. Due to extended phase match (EPM) realized in our crystal near 792 nm, the central wavelength of the photon generated in present setup can be continuously tuned from 1540 nm to 1600 nm, covering the C+L telecom bands. The tuning range can excess 100 nm if a temperature controller with wider tuning range and a more efficient single photon detector (like the one used in [18]) are available in our Lab. Bell-type interferences are measured at different pump wavelength. The S parameter value of CHSH inequality with accidental coincidences subtracted is in the range between 2.60±0.04 to 2.72±0.07, which clearly demonstrates the high entanglement of the source over the whole tunable range. We believe our source could definitely improve the transmission capacity of a single communication channel in the field of quantum communication if it combines with WDM technique.

The text is organized as follows: we simply explain the reason why we can realize such a broadband tuning ability and also investigate the temperature tuning property of the crystal by performing the second harmonic generation (SHG) experiment in section 2. Then we describe the experimental layout of our photon source in section 3. Details of the tuning ability and the evaluation of entanglement are showed in section 4, factors that defect the entanglement of the present setup are also discussed there. Then we come to the conclusion section. The main characters of our source are summarized, some possible improvements and potential applications are also discussed there.

## 2. Extended phase match and temperature tuning property of the crystal

Here we consider the generation of a frequency-degenerated photon pair, in which the frequencies of the signal and idler photon equal to the half of the pump frequency. The phase-match condition for the degenerated collinear down-conversion in a type-II PPKTP is

$$\Delta k \equiv k_p(\omega_p) - k_s(\omega_p/2) - k_i(\omega_p/2) + 2\pi/\Lambda = 0$$, where $k_{p,s,i}$ is wave

vector of the pump ($p$), signal ($s$) and idler ($i$) respectively, $\Delta k$ is phase mismatch, and $\Lambda$ is the poling period of the crystal. If the pump wavelength changes, $\Delta k$ is no longer zero for degenerated SPDC outputs. But this can be rectified if we operate at certain pump wavelength where the phase mismatch's first order frequency derivative $\Delta k' = \partial(\Delta k)/\partial\omega$ is also zero, which yields $k'(\omega_p) = \left(k'_s(\omega_p/2) + k'_i(\omega_p/2)\right)/2$. The phase match of satisfying both $\Delta k = 0$ and $\Delta k' = 0$ is called extended phase match (EPM) [30, 31]. Thanks to EPM, we could obtain degenerated SPDC outputs during wide pump wavelength range by simply tuning the temperature of the crystal in a small range. Such EPM can't be achieved when the pump wavelength of the SPDC is far away from this certain wavelength. In this case, dramatically wide temperature tuning range of the crystal is necessary in order to achieving a broadband tuning ability, obviously this is not practical. The EPM wavelength of our crystal is near 792 nm, which is calculated from the dispersion equation given in [32, 33].

To achieve degenerated SPDC outputs at different pump wavelengths, we need the exact phase match temperatures. These temperatures are measured using SHG process. In follow text, the temperature tuning property is measured and shown. The crystal we use is bought from Raical Crystals, which is x-cut with a dimension of 1×2×10 $mm^3$. The crystal is periodically poled with a periodicity of 46.2 $\mu m$ to get first-order quasi phase match (QPM) for SPDC at the pump wavelength of 780 nm and signal and idler wavelengths at 1560 $nm$. Both end faces of the crystal are anti-reflect coated at wavelengths of 780 $nm$ and 1560 $nm$. The crystal is mounted on a three-dimensional translation stage, and aligned by four axis tilting. The temperature of the crystal is controlled by a semiconductor Peltier in the range of -10℃～70 ℃ with a stability of ±0.01℃.

The SHG is performed using a commercial DFB diode laser (DL prodesign, Toptica). The wavelength of output laser can be continuously tuned from 1529 nm to 1595 nm. We measure the SHG power as the function of temperature of crystal at different wavelengths, results are shown in figure 1(a). We plot SHG power against the temperature of crystal at wavelengths of 1545 nm, 1550 nm, 1555 nm and 1560 nm respectively in figure 1(a). According to the Gaussian fit to the experimental data in figure 1(a), we obtain a temperature bandwidth of about 80℃. We also plot the optimal phase match temperature as the function of wavelength in figure 1(b). The square points in figure1(b) are experimental measured data, and the solid lines are calculated using dispersion equations given in Refs. 32 and 33 and temperature coefficient given in Ref. 34. There is a high agreement between experimental data and theoretical calculation. We find that the degenerate output temperature has a turning point near 1584 nm, and the temperature tuning curve is symmetry with respective to this point. It is this turning point that is responsible for the broadband tuning ability of our source. If the source operates far away from this point, the temperature never turns back when you tune the wavelength of the pump along one direction. After having the phase match temperature at different wavelengths, we could make the spectrums of the signal and idler obtained by SPDC overlapped by tuning the temperature of the crystal to a certain value in a wide pump wavelength range.

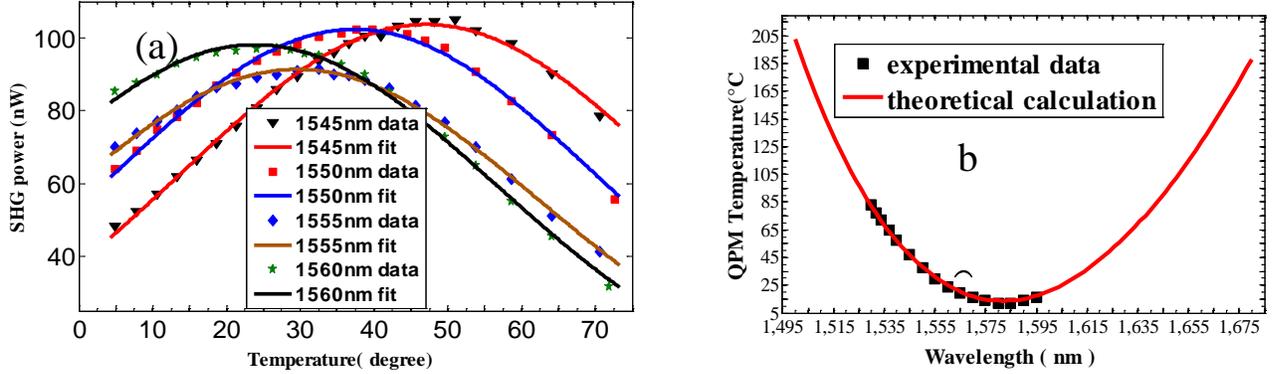

**Figure 1.** (a) SHG power as function of temperature at wavelengths of 1545 nm, 1550 nm, 1555 nm and 1560 nm. (b) Phase match temperature at different wavelength. Squared points are data from SHG experiments, solid cures are theoretical calculations.

## 3. Experimental Scheme

The layout of our continuously tunable source is depicted in figure 2. The pump light beam is from a continuous wave (CW) Ti: Sapphire laser (Coherent MBR110), its wavelength can be continuously tuned from 756 nm to 870 nm. The polarization of the pump beam is controlled with a half wave plate. The beam is focused into the PPKTP crystal by lens L1. The group delay inside the PPKTP crystal is compensated by a 5-mm long KTP crystal with its optical axis rotated by 90 degree relative to the PPKTP crystal. The strong pump beam is filtered using a dichromatic mirror and a long-pass filter. After that, the signal and idler photons are collected using a single mode fiber (SMF). Then we use all fiber components to do Bell-type interference. The Bell-type interference part contains a 50/50 fiber beam splitter and two fiber polarization rotators and polarizers (FPR&P). The output of FPR&P1 is connected to a single-photon detector InGaAs avalanche photodiode 1(APD1) directly, and the output of FPR&P2 is optically delayed using a 200-m long SMF before connecting to APD2. APD1 is running at internal trigger mode with a maximum trigger rate of 10 MHz, the output of APD1 is electrically delayed using a digital delay generator DG535 ( Stanford Instr.). The electrical signal from DG535 is used to trigger APD2. The output of APD2 is connected to a counter to accumulate the coincidence counts.

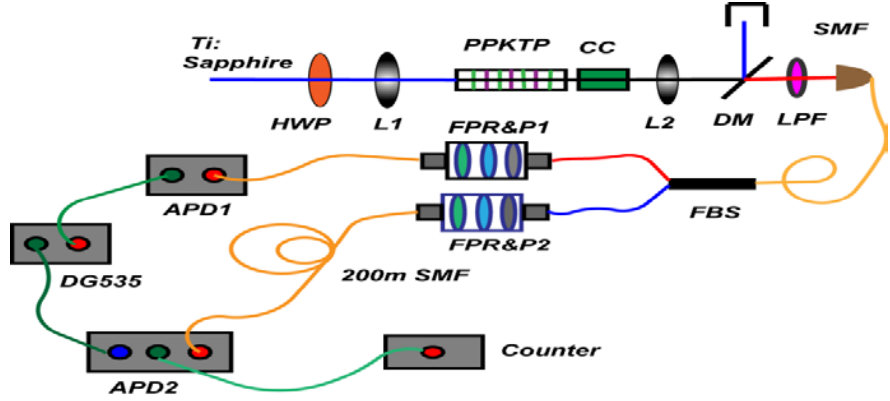

**Figure 2.** The layout of the experiment. HWP: half wave plate; L1, L2: lens; PPKTP: periodically-poled KTiOPO$_4$ crystal; CC: KTP compensate crystal; DM: dichromatic mirror; LPF: long pass filter; SMF: single mode fiber; FBS: fiber beam splitter; FPR&P1, 2: fiber polarization rotator and polarizer; APD1, 2: InGaAs avalanche photodiode; DG535: digital delay generator.

## 4. Tuning property and evaluation of source

To demonstrate the tuning ability of the present setup, the wavelength of the pump beam is scanned from 770 nm to 800 nm with a step of 5 nm, the temperature of the crystal is tuned at the same time to get degenerated SPDC output. At each pump wavelength, we perform the Bell-type interference measurement to characterize the quality of the source. At certain pump wavelength, the spectral difference of down-converted signal and idler photons is removed by tuning the temperature of the crystal to the corresponding value obtained in the process of the SHG, and the temporal distinguishability is eliminated by the compensation crystal. After being coupled into the SMF, the signal and idler photons are also spatially indistinguishable. The post selected state [8, 35] after the fiber beam splitter is $|\phi\rangle = 1/\sqrt{2}(|V\rangle_R|H\rangle_T + |H\rangle_R|V\rangle_T)$, where R and T represent the reflecting and transmitting of signal and idler photons respectively.

We measure the polarization interference of the Bell-state in the 0/90 degree and +/-45 degree bases respectively. The results are presented in figure 3. Figures 3(a) and (b) are experimental data obtained at pump wavelength of 770.266 nm and 800.139 nm. The interference visibility is defined as $V = (C_{\max} - C_{\min})/(C_{\max} + C_{\min})$, where $C_{max(min)}$ represents the maximum (minimum) coincidences. In figure 3(a), the raw (net) visibilities of the 0/90 degree and +/-45 degree bases are 89.83±2.94% (98.54%) and 79.89±2.86% (89.07%) respectively and the corresponding visibilities in figure 3 (b) are 87.60±3.03% (97.14%) and 87.26±2.02% (98.96%). As is well known that a visibility greater than 71% implies a violation of CHSH inequality [36], the experimentally obtained high visibilities are sufficient for the violation of Bell-inequality, and imply a high polarization entanglement of our source. The experimental data at other pump wavelengths are not given here, but the visibilities of them obtained from experimental data are depicted in figure 3(c). Figure 3(c) shows that the raw visibility is near 80% in +/-45 degree base, and is near 90% in the 0/90 degree base at any pump wavelength. The visibility with accidental coincidences subtracted is higher than the respective raw visibility. We also investigate the impact of the pump power to the ratio of the maximum coincidences to the minimum coincidences in two bases. The pump wavelength is fixed at 780.457 nm in this case. The results are showed in figure 3 (d). It is obvious that the ratio decreases with the increase of pump power. The ratio is much more sensitive in the 0/90 degree base than that in +/-45 degree base, the imperfect overlap of the signal and the idler at the polarizer prevents increase of the ratio in the+/-45 degree base. The reduction of the ratio is a result of accidental coincidences caused by multi-photon emission, so the raw visibility could be improved if a lower power pump is used.

In order to demonstrate further the entanglement between the photons in a pair, we check the CHSH inequality. It is well known that CHSH inequality $|S| \leq 2$ holds if there is no entanglement. On the contrary, it is violated if there is entanglement. Where S is defined as [36, 37]

$$S = E(\theta_1, \theta_2) + E(\theta_1', \theta_2) + E(\theta_1, \theta_2') + E(\theta_1', \theta_2'), \quad (1)$$

and $E(\theta_1, \theta_2)$ is given by

$$\frac{C(\theta_1, \theta_2) + C(\theta_1^\perp, \theta_2^\perp) - C(\theta_1, \theta_2^\perp) - C(\theta_1^\perp, \theta_2)}{C(\theta_1, \theta_2) + C(\theta_1^\perp, \theta_2^\perp) + C(\theta_1, \theta_2^\perp) + C(\theta_1^\perp, \theta_2)}. \quad (2)$$

Where, $C(\theta_1, \theta_2)$ is coincidence counts at different settings of the two polarizer.

The settings we used in our experiments are $\theta_1 = -22.5^o$, $\theta_1^\perp = 67.5^o$; $\theta_1' = 22.5^o$, $\theta_1'^\perp = 112.5^o$; and $\theta_2 = -45^o$, $\theta_2^\perp = 45^o$; $\theta_2' = 0^o$, $\theta_2'^\perp = 90^o$. We calculate the S parameter of the CHSH inequality using the data we measure and give the amount of standard deviations. The results are showed in figure 3(e), (f). The maximum (minimum) S parameter value with accidental coincidences subtracted showed in figure 3(e) is 2.72±0.07 at 1600 nm (2.60±0.04 at 1570 nm), which clearly demonstrates the high entanglement of the source at different pump wavelengths. In figure 3(f), the maximum (minimum) standard deviation is 21.4 at 1560 nm (8.0 at 1540 nm).

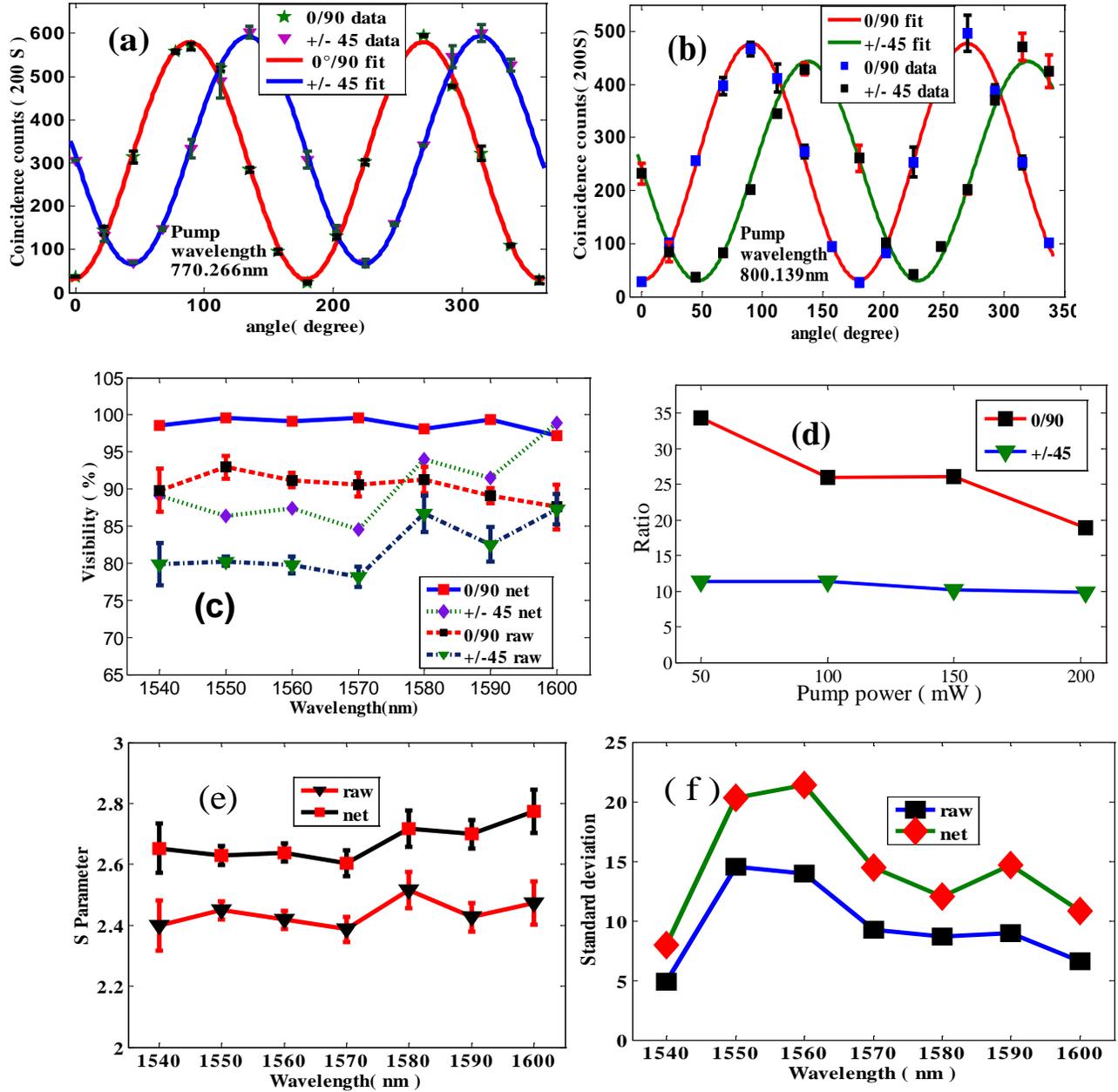

**Figure 3.** (a),(b) Bell-type interference curves. Coincidences counts in 200 seconds as a function of the angle of polarizer 1, the angle of polarizer 2 is fixed at $0^0$ or $45^0$, the pump wavelengths are 770.266 nm and 800.139 nm in (a), (b) respectively; (c) raw (net) visibility in $0^0/90^0$ and +/-$45^0$ bases at different wavelength of the down-converted photons; (d) ratio

of maximum coincidences to minimum coincidences as a function of pump power in different bases;(e) S parameters of the CHSH inequality at different down-converted wavelength, the black line is the value with accidental coincidences subtracted, red line is the values with the accidental coincidences;(f) standard deviations at different down-converted wavelengths.

Before each measurement data points, the polarization in the fiber is calibrated using the diode laser (The same as the laser used in the SHG part). Some other parameters of each data points are list in Table 1.

Table 1. Parameters in the experiment

| Pump wavelength(nm) | Pump power(mW) | Crystal temperature(℃) | Single count of APD1($S^{-1}$) |
|---|---|---|---|
| 770.266 | 190 | 58.0 | 2000±100 |
| 775.305 | 190 | 38.2 | 2400±100 |
| 780.457 | 205 | 23.1 | 2700±100 |
| 785.171 | 202 | 16.0 | 2400±100 |
| 790.412 | 192 | 12.1 | 2100±100 |
| 795.244 | 196 | 14.7 | 1900±100 |
| 800.139 | 190 | 18.5 | 1600±100 |

From table 1, we conclude that the detection efficiency of the InGaAs avalanche photodiode decreases very fast in the longer wavelength. The average dark counts (10MHz trigger rate, 5 ns detection window) are 310 $s^{-1}$. A lower detection efficiency means longer accumulation time, this limits further increase of pump wavelength using the present single photon detector. The pump wavelength is also limited by the range of the temperature controller. If the controlled temperature can be as high as 150℃, then the wavelength of the pump can be tuned from 760 nm to 830 nm, this implies a tuning range about 140 nm of down-converted fields.

The bandwidth of the down-converted photons is 2 nm, which is calculated from the phase mismatch $\Delta k = 2\pi[\frac{n_y(\lambda_p)}{\lambda_p} - \frac{n_y(\lambda_s)}{\lambda_s} - \frac{n_z(\lambda_i)}{\lambda_i} + \frac{1}{\Lambda}]$ using the dispersion formula given in [29, 30]. The photon collecting efficiency in our experiment is 0.2212 (6.55 dB loss, the loss of the filter elements before coupling into SMF is 1.6 dB, and the fiber coupling loss is 4.95d B.) The insertion losses of the two FPR&P are 0.31 dB and 0.66 dB, respectively. Accounting for all loss factors listed above and the detection efficiency of detector (8%), we estimate a photon pair generation rate of 1.63×10$^4$(s·mW·nm)$^{-1}$.

Though the visibility showed in figure 3(c) is high enough to demonstrate the entanglement of the photons in a pair, there are still some rooms for further improvement. As the temperature of the compensation crystal is not stabilized in present experiment, the temporal distinguishbility of the two photons is not removed completely due to the temperature differences between the two crystals. Thus the visibility will be better if the temperature of the compensation crystal is also controlled precisely. Another factor that defects the visibility is the polarization dispersion and drifting in the fiber. It takes hours to measure data at a fixed pump wavelength. The fluctuation of the environment temperature will lead to polarization drifting of the photons in the fiber. A high collecting efficiency of the down-converted photons should be very promising for practical using of our source. This is possible by using low loss filter elements and proper designing of zoom-lens (in Ref. 18, a single mode fiber coupling efficiency of 80% is realized).

## V. Conclusion

In summary, an ultra-broadband continuously-tunable, high polarization-entangled photon-pair source in the C+L telecom bands is demonstrated. A tuning range of 60 nm of the down-converted photons is achieved using the present setup, and the range over 100 nm is expected. The photon generating rate and collecting efficiency can improve greatly if a waveguide PPKTP is used. Our source could be used to burst the transmission capacity of a single communication channel in the field of quantum communication.

## Acknowledgements

We thank Dr. Wei Chen for kindly loan us single photon detector and for other technique support. This work was supported by the National Natural Science Foundation of China (Grant Nos., 11174271, 61275115), the National Fundamental Research Program of China (Grant No. 2011CB00200), and the Innovation fund from CAS, Program for NCET.

## References and links